# Relationship between Thermodynamic Driving Force and One-Way Fluxes in Reversible Processes

Daniel A. Beard
Department of Physiology
Medical College of Wisconsin

Hong Qian
Department of Applied Mathematics
University of Washington

November 18, 2006

**Chemical reaction systems operating in nonequilibrium open-system states arise in a great number of contexts, including the study of living organisms, in which chemical reactions, in general, are far from equilibrium. Here we introduce a theorem that relates forward and reverse fluxes and free energy for any chemical process operating in a steady state. This relationship, which is a generalization of equilibrium conditions to the case of a chemical process occurring in a nonequilibrium steady state, provides a novel equivalent definition for chemical reaction free energy. In addition, it is shown that previously unrelated theories introduced by Ussing and Hodgkin and Huxley for transport of ions across membranes, Hill for catalytic cycle fluxes, and Crooks for entropy production in microscopically reversible systems, are united in a common framework based on this relationship.**

## Introduction

For a reaction occurring in an isothermal and isobaric system the chemical driving force $\Delta G$—the Gibbs free energy difference—characterizes how far a chemical reaction is away from equilibrium. If we take a simple bimolecular reaction in a dilute solution

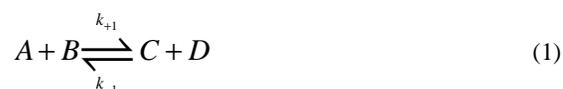

$$A + B \underset{k_{-1}}{\overset{k_{+1}}{\rightleftharpoons}} C + D \qquad (1)$$

as an example, then $\Delta G$ is related to the concentrations of the reactants and products, as well as the equilibrium constant $K_{eq}$, through the well-known thermodynamic equation

$$\Delta G = -RT \ln \left( [A][B] K_{eq}/[C][D] \right). \quad (2)$$

If we further assume that the law of mass action governs the reaction's kinetics, then the forward and reverse reaction fluxes and equilibrium constant are

$$J^+ = k_{+1}[A][B], \quad J^- = k_{-1}[C][D], \quad K_{eq} = k_{+1}/k_{-1}, \quad (3)$$

where $k_{+1}$ and $k_{-1}$ are constants that do not depend on the concentrations. Combining Equations (2) and (3) yields

$$\Delta G = -RT \ln \left( J^+/J^- \right). \quad (4)$$

Expressing $\Delta G$ in terms of Equation (4) has many advantages: It is apparent that if $\Delta G = 0$, then $J^+ = J^-$. This equilibrium relationship is required by the principle of detailed balance, which states that at equilibrium the forward and reverse fluxes are equal for all existing independent mechanisms for the reaction $A \rightleftharpoons B$ [10]. Furthermore, Equation (4) can be generalized to many other situations. For example, for reversible enzyme reactions governed by Michaelis-Menten kinetics, although both $J^+$ and $J^-$ are complex, nonlinear functions of reactant and substrate concentrations, Equation (4) still holds true.

Another nontrivial example of Equation (4) that arises in cycle kinetics in unimolecular systems is due to T.L. Hill [3-6]. As in the example above, the law of mass action is assumed in all of Hill's work. The novelty of this note is to show a wide range of validity of Equation (4) based solely on conservation of mass, without invoking any assumptions of rate laws such as Equation (3). Hence, Equation (4) is in fact a fundamental relation for any chemical process operating in steady, open-system state.

The relation is also intimately related to the fluctuation theorem [1,2,14,17]. However, the most significant insight from the present work is that the relation between one-way-fluxes and $\Delta G$ can be established without any supposition on the dynamics of a system.

## Flux and Free Energy in a Nonequilibrium Open System

For the reaction $A \rightleftharpoons B$, the Gibbs free energy change per mole of molecules that transform from state A to state B is expressed

$$\Delta G = \Delta G^o + RT \ln(N_B/N_A), \qquad (5)$$

where $N_A$ and $N_B$ are the number of molecules in states A and B, respectively. In equilibrium, the ratio $N_B/N_A$ is equal to $e^{-\Delta G^o/RT} = K_{eq}$ and the net reaction flux is $J = J^+ - J^- = 0$, where $J^+$ and $J^-$ are the forward and reverse reaction fluxes, respectively. When $\Delta G < 0$, the net flux $J$ from A to B is positive.

To determine how flux and free energy are related for systems not in equilibrium we consider, without loss of generality, the case where $N_B/N_A < K_{eq}$ and $J > 0$. In a nonequilibrium steady state $N_A$ and $N_B$ are held constant by pumping A molecules into the system, and pumping B molecules out of the system, at the steady-state flux $J$.

Next imagine that we are able to place a label on each molecule that converts from state B to state A. These particles we denote by A*. Apart from the label, A* molecules are identical in every way to unlabeled A molecules in this thought experiment. In addition, imagine that A* molecules lose their label when they convert to B molecules. Thus if we continue to pump A and B molecules into and out of the system at the constant flux $J$, then a steady state will be reached for which $N_{A*}$, the number of labeled molecules in state A*, is less than or equal to $N_A$, the total number of labeled plus unlabeled molecules in state A.

The steady state is reached when the rate of conversion of labeled A* molecules into state B is equal to the rate of conversion from B to A*. Since there is no transport of A* into or out of the system, then in the steady state the $N_{A*}$ molecules in state A* will be in equilibrium with the $N_B$ molecules in state B: $N_B/N_{A*} = K_{eq}$. Mass conservation requires that the forward flux of A* → B equal the reverse flux of B → A*, or $J^+ \frac{N_{A*}}{N_A} = J^-$. Combining these equations, we have:

$$\frac{J^+}{J^-} = \frac{N_B}{N_A} K_{eq}. \qquad (6)$$

This relationships hold for a reaction operating in any steady state, including thermodynamic equilibrium. In equilibrium, $J^+ = J^-$, and

$$\frac{N_B}{N_A} K_{eq} = 1. \tag{7}$$

Thus it is trivial that Equation (6) holds in equilibrium. The more interesting case is a nonequilibrium steady state for which Equations (5) and (6) yield Equation (4). Therefore Equation (4) is a condition that does not depend on the details of the kinetic reaction mechanism that is operating in a particular system. In addition, the above proof is easily generalized to apply to multimolecular (non-uni-unimolecular) chemical reactions or any spontaneous process transforming or transporting mass from one state to another. Therefore Equation (4) represents a fundamental property of any chemical process.

## Relationship to other Theories

### Hill Equation for Catalytic Cycles

For the case of a catalytic cycle with $J^+/J^-$ equal to the ratio of the forward-to-reverse cycle flux and $\Delta G$ equal to the thermodynamic driving force for the cycle, Equation (4) is identical to the relationship introduced by Hill [3,7,6] and proved by Kohler and Vollmerhaus [9] and by Qian et al. [13] for cycles in Markov systems. (See Equations (3.7) and (7.8) in [6].) Therefore the relationship between $J^+/J^-$ and $\Delta G$ introduced by Hill for linear cycle kinetics is a special case of Equation (4).

As specific example, consider the well known Michaelis-Menten enzyme mechanism:

$$A + E \underset{k_{-1}}{\overset{k_{+1}}{\rightleftharpoons}} C \underset{k_{-2}}{\overset{k_{+2}}{\rightleftharpoons}} B + E, \tag{8}$$

in which E is an enzyme involved in converting substrate A into product B. The steady-state flux through this mechanism is

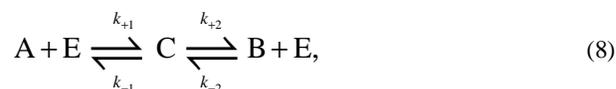

$$J_{MM}(a,b) = J^+ - J^- = \frac{k_f a - k_r b}{1 + a/K_a + b/K_b}, \tag{9}$$

where $E_o$ is the total enzyme concentration, $a = [A]$, $b = [B]$, $k_f = E_o k_{+1} k_{+2}/(k_{-1} + k_{+2})$, $k_r = E_o k_{-1} k_{-2}/(k_{-1} + k_{+2})$, $K_a = (k_{-1} + k_{+2})/k_{+1}$, and $K_b = (k_{-1} + k_{+2})/k_{-2}$. Identifying $J^+$ as the positive term and $J^-$ as the negative term in Equation (9), it is straightforward to verify that $J^+$ and $J^-$ satisfy Equation (4), where $\Delta G^o = -RT \ln(k_{+1} k_{+2}/k_{-1} k_{-2})$.

## Crooks Fluctuation Theorem

Again, we consider a system made up of molecules that can transition between two states: $A \rightleftharpoons B$. If $\Gamma^+$ is the mean forward transition rate (number of transitions per unit time) for which the system is driven in the forward direction, then the probability of $n$ forward transitions in a finite time period $\tau$ is given by the Poisson distribution: $P_+(n) = (\tau \Gamma^+)^n e^{-\tau \Gamma^+} / n!$. Likewise, the probability of $m$ reverse transitions is $P_-(m) = (\tau \Gamma^-)^m e^{-\tau \Gamma^-} / m!$, where $\Gamma^-$ is the reverse transition rate. Hence, the probability of net forward turnover $l = m - n$ is given by

$$P_F(l) = \sum_{n=l}^{\infty} P_+(n) \cdot P_-(n-l) = (\tau \Gamma^-)^{-l} \sum_{n=l}^{\infty} \frac{(\tau^2 \Gamma^+ \Gamma^-)^n e^{-\tau(\Gamma^+ - \Gamma^-)}}{n!(n-l)!} \tag{10}$$

and the probability $l$ net reverse turnovers is

$$P_R(l) = \sum_{m=l}^{\infty} P_+(m-l) \cdot P_-(m) = (\tau \Gamma^+)^{-l} \sum_{m=l}^{\infty} \frac{(\tau^2 \Gamma^+ \Gamma^-)^m e^{-\tau(\Gamma^+ - \Gamma^-)}}{(m-l)!m!} \tag{11}$$

Associated with net $l$ forward turnovers is the entropy production (e.p.) of $-(l\Delta G)$; the net $l$ reverse turnovers have an e.p. of $+(l\Delta G)$. Therefore, the ratio of the probability of e.p. $= \sigma$ to the probability of e.p. $= -\sigma$, within a finite time interval, is

$$\frac{\Pr(\text{e.p.} = +\sigma)}{\Pr(\text{e.p.} = -\sigma)} = \frac{P_F(l = +\sigma/\Delta G)}{P_R(l = -\sigma/\Delta G)} = \left(\frac{\Gamma^+}{\Gamma^-}\right)^{\frac{-\sigma}{\Delta G}} \tag{12}$$

Connecting this result with Equation (4), we have

$$\frac{\Pr(\text{e.p.} = +\sigma)}{\Pr(\text{e.p.} = -\sigma)} = e^{\frac{\sigma}{RT}}, \tag{13}$$

which is known as the Crooks fluctuation theorem [1].

## Ussing Flux Ratio

When a charged species is transported across a biological membrane, the ionic flux is influenced by any electrostatic potential difference that exists across the membrane. Using the convention that the electrostatic potential across a cell is measured as the outside potential subtracted from inside potential,

Equation (4) is expressed

$$\frac{J^{out}}{J^{in}} = \exp\left(\frac{-\Delta G}{RT}\right) = \frac{c_i}{c_o}\exp\left(\frac{zF\Delta\Psi}{RT}\right) \qquad (14)$$

for passive transport of a single ion across a cell membrane. The concentrations $c_i$ and $c_o$ in Equation (14) denote the concentrations of the ion on the inside and outside of the cell, respectively; $z$ is the valence number of the ion, $F$ is Faraday's constant, and $\Delta\Psi$ is the electrostatic potential. The flux ratio in form of Equation (14) and applied to single-ion transport was introduced in 1949 by Ussing [19] and is known as the Ussing flux ratio. Based on the assumption that $J^{out}$ is independent of $c_o$ and that $J^{in}$ is independent of $c_i$, Hodgkin and Huxley derived the same expression in 1952 [8].

The current work shows that the theory of Ussing and Hodgkin and Huxley is a special case of Equation (4). In addition to single-ion channel fluxes for which the Ussing flux ratio has been developed and applied, the flux ratio of Equation (4) applies to all active and passive transport processes as well as multiple-ion transporters.

## Additional Consequences

Net Flux for Nearly Irreversible Reactions is Proportional to Reverse Flux

We can study nearly irreversible systems based on Equation (4). The net flux through a chemical process is $J = J^+ - J^-$; thus, $e^{-\Delta G/RT} = J/J^- + 1$, which leads to the approximation

$$J = J^- e^{-\Delta G/RT} \qquad (15)$$

for nearly irreversible reactions ($J \gg J^-$). Thus the net flux through an enzyme in a reaction operating far from equilibrium is proportional to the reverse flux.

For the quasi-steady approximation of Equation (9) the reverse flux is $J^- = k_r b/(1 + a/K_a + b/K_b)$; thus for a nearly irreversible reaction

$$J = J^- K_{eq}\frac{a}{b} = \frac{k_f a}{1 + a/K_a + b/K_b}, \qquad (16)$$

which is the expression that we would arrive at by setting $k_r = 0$ in Equation (9). Note that the usual irreversible Michaelis-Menten equation derives from the assumption that $k_{-2} = 0$, which results in

$$J = \frac{k_f a}{1 + a/K_a}. \tag{17}$$

This analysis illustrates that the assumption $k_{-2} = 0$ is a special case of the irreversible single-substrate enzyme. Equation (16) is the general approximation for the case of $|\Delta G/RT| \gg 1$, where $k_{-2}$ may be finite.

## Net Flux for Highly Reversible Reactions is Proportional to Reverse Flux

Near equilibrium (for $|\Delta G| \ll RT$) the flux can be approximated as linearly proportional to the thermodynamic driving force: $J = -X \Delta G$, where $X$ is the Onsager coefficient [11,12]. When the near-equilibrium approximation $|\Delta G| \ll RT$ holds, the flux ratio $J^+/J^-$ is approximately equal to 1. In this case Equation (4) is approximated

$$\Delta G = -RT(J^+/J^- - 1) = \frac{RT}{J^-}(J^+ - J^-). \tag{18}$$

From this expression, we have

$$J = -\frac{J^-}{RT} \Delta G. \tag{19}$$

Therefore for highly reversible systems, the net flux is proportional to the reverse flux times the thermodynamic driving force; the Onsager coefficient is equal to $J^-/RT$.

## Application to Transport Processes

In addition to application to chemical reactions, Equation (4) is directly applied to transport processes. For example, one-dimensional transport of particles in a complex medium is governed by a Fokker-Planck equation with spatially dependent diffusion coefficient $D(x)$ and potential function $u(x)$ [18]:

$$\frac{\partial c(x,t)}{\partial t} = \frac{\partial}{\partial x}\left[\frac{D}{RT} c \frac{\partial u}{\partial x} + D \frac{\partial c}{\partial x}\right], \tag{20}$$

over the domain $0 \leq x \leq 1$. The steady-state transport flux predicted by this equation is

$$J = J^+ - J^- = -\frac{D}{RT}c\frac{\partial u}{\partial x} - D\frac{\partial c}{\partial x} = \left(c_0 e^{u(0)/RT} - c_1 e^{u(1)/RT}\right)\left[\int_0^1 e^{u(y)/RT}\frac{dy}{D(y)}\right]^{-1}, \quad (21)$$

where $c_0$ and $c_1$ are the concentrations of the two reservoirs at $x=0$ and $x=1$. Recognizing that $\Delta G = -\{u(1) - u(0) + RT\ln(c_1/c_0)\}$ for this system, we have Equation (4).

## Exchange of Isotope Labels

A variety of isotope labeling methods are used to determine in vivo metabolic fluxes. In some cases, it is possible to estimate not only the net flux of a given reaction, but also the forward and reverse rate at which an isotope label exchanges between species involved in a chemical reaction [20]. Consider as examples the enzyme-mediated catalysis schemes for the reaction $A \rightleftharpoons B$ illustrated in Figure 1.

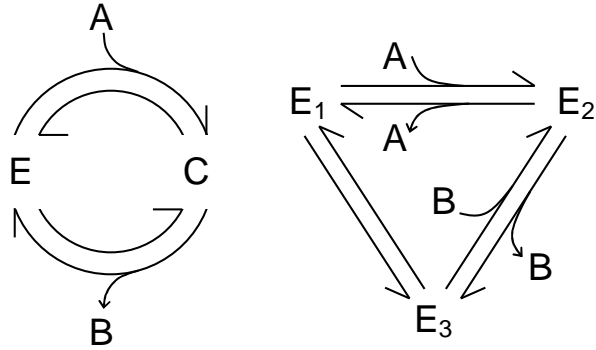

Figure 1: Example enzyme mechanisms for the reaction $A \rightleftharpoons B$. The left panel illustrates the Michaelis-Menten scheme of Equation (8), in which enzyme binds to substrate A, forming a complex C. The product B reversibly dissociates from the complex C, forming unbound enzyme E. The right panel illustrates a more complex mechanism involving three enzyme states $E_1$, $E_2$, and $E_3$. Enzyme kinetic theory assumes that the state transitions follow mass-action kinetics, as described by Equation (8) for the left panel and Equation (23) for the right panel.

For the reversible Michaelis-Menten example of the left panel, which is described by Equations (8) and (9), the exchange flux ratio—the rate at which a label on A molecules is transferred to B molecules divided by the rate at which a label on B molecules is transferred to A molecules is given by Equation (4),

$$\left(\frac{J^+}{J^-}\right)_{exchange} = \exp\left(\frac{-\Delta G_{AB}}{RT}\right) = K_{eq} \cdot \frac{a}{b}, \quad (22)$$

where $\Delta G_{AB}$ and $K_{eq}$ are the Gibbs free energy and equilibrium constant for the reaction.

For the mechanism illustrated in the right panel of the figure, the exchange flux ratio takes a slightly different form. This enzyme mechanism has the following elementary steps:

$$A + E_1 \underset{k_{-1}}{\overset{k_{+1}}{\rightleftharpoons}} E_2$$

$$E_2 \underset{k_{-2}}{\overset{k_{+2}}{\rightleftharpoons}} E_3 + B \quad (23)$$

$$E_3 \underset{k_{-3}}{\overset{k_{+3}}{\rightleftharpoons}} E_1$$

where $K_{eq} = (k_{+1} \cdot k_{+2} \cdot k_{+3})/(k_{-1} \cdot k_{-2} \cdot k_{-3})$. The overall flux ratio for the enzyme cycle $E_1 + A \rightleftharpoons E_1 + B$, has the same form as Equation (22)

$$\left(\frac{J^+}{J^-}\right)_{cycle} = \exp\left(\frac{-\Delta G_{AB}}{RT}\right) = K_{eq} \cdot \frac{a}{b}. \quad (24)$$

The exchange flux ratio follows from applying Equation (4) to the reaction $E_1 + A \rightleftharpoons E_3 + B$:

$$\left(\frac{J^+}{J^-}\right)_{exchange} = K_{eq} \frac{a}{b} \cdot \frac{k_{-3}[E_1]}{k_{+3}[E_3]} \leq \frac{-\Delta G_{AB}}{RT}. \quad (26)$$

The inequality in Equation (25) assumes that the net flux is positive and $k_{-3}[E_1] \leq k_{+3}[E_3]$.

## Summary

In summary, we have demonstrated that Equation (4) is a fundamental condition that is satisfied by any chemical process operating in a steady state. This equation is a generalization of the well known equilibrium conditions $\Delta G = 0$ and $J^- = J^+$ to the case of a chemical process occurring in a non-equilibrium steady state, such as a chemical reaction in an open system [15,16]. It provides a novel equivalent definition for the reaction free energy, or thermodynamic driving force. Based on this relationship, related theories of Ussing [19] and Hodgkin and Huxley [8] for ionic transport across membranes, Hill [3-7] for enzyme cycle kinetics, and Crooks [1] for entropy production and work done by microscopically reversible systems, are united in a common framework.

We thank J.B. Bassingthwaighte and A. Cornish-Bowden, for valuable comments on the manuscript. Discussions regarding the Ussing flux ratio with James Sneyd were of great value, as were discussions with Wolfgang Wierchert on labeled isotope exchange. This work was supported by NIH grant GM0680610.